\documentclass[]{elsart5p}
\usepackage{graphicx}
\usepackage{latexsym}   
\journal{Physica A}

\begin{document}

\begin{frontmatter}

\title{Continuous growth models in terms of generalized logarithm and exponential functions}

\author[Filo]{Alexandre Souto Martinez\corauthref{cor}},
\corauth[cor]{Corresponding author.}
\ead{asmartinez@ffclrp.usp.br}
\author[Filo]{Rodrigo Silva Gonz\'alez},  
\author[Filo,Barao]{C\'esar Augusto Sangaletti Ter\c{c}ariol} 
 
\address[Filo]{Faculdade de Filosofia, Ci\^encias e Letras de Ribeir\~ao Preto (FFCLRP) \\
               Universidade de S\~ao Paulo (USP) \\
               Av.~Bandeirantes, 3900 \\
               14040-901  Ribeir\~ao Preto, SP, Brazil}
\address[Barao]{Centro Universit\'ario Bar\~ao de Mau\'a \\
                Rua Ramos de Azevedo, 423 \\ 
                14090-180, Ribeir\~ao Preto, SP, Brazil}

\newpage

\begin{abstract}
Consider the one-parameter generalizations of the logarithmic and exponential functions which are obtained from the integration of non-symmetrical hyperboles. 
These generalizations coincide to the one obtained in the context of non-extensive thermostatistics. 
We show that these functions are suitable to describe and unify the great majority of continuous growth models, which we briefly review. 
Physical interpretation to the generalization function parameter is given for the Richards' model, which has an underlying microscopic model to justify it.  
\end{abstract}

\begin{keyword}
population dynamics \sep 
generalized logarithmic and exponential functions \sep 
growth models \sep 
Richard's model \sep
Marusic and Bajzer's model \sep 
Tsoularis and Wallace's model  
\PACS  87.23.Cc \sep  %
       89.75.-k
\end{keyword}

\end{frontmatter}

\newpage

\section{Introduction}

The convenience of generalizing the logarithmic and exponential functions has attracted the attention of researchers in the last years. 
One-parameter generalizations of logarithmic and exponential functions have been proposed in the several contexts~\cite{tsallis_qm,nivanen_2003,borges_2004,kalogeropoulos_2005,kaniadakis_2001,PhysRevE.66.056125,abe_1997}. 
More sophisticated generalizations have also been proposed such as a two and three-parameter~\cite{kaniadakis:046128,kaniadakis:036108}, which includes all the others generalized logarithms as special cases. 
Examples of the convenience of these generalizations have been seen in different fields~\cite{takahashi_2_2008,takahashi_1_2008,takahashi_2_2007,takahashi_1_2007,cajueiro_2006,takahashi_3_2007,anteneodo:1:2002,holanda:2004,albuquerque:2000}. 

From an initial population size $n_0$, at an initial time $t_0$, growth models are used to predict the size $n(t)$ of population at an arbitrary time $t > t_0$. 
The most important parameters are the intrinsic growth rate $r$ and carrying capacity $n_{\infty} = n(t \rightarrow \infty)$. 
Nevertheless frequently additional parameters are needed to adjust the model to a given experimental data. 
Up to date, a comprehensive description of continuous growth models is given by the Tsoularis and Wallace model~\cite{tsoularis_2002}, which generalizes the logistic-like models~\cite{murray,boyce_diprima,richards_1959,spencer_1966,blumberg_1968,turner_1976,marusic_1993}. 

The main objective of this paper is to show that the one-parameter generalization of the logarithmic and exponential functions proposed~\cite{tsallis_qm} is suitable to describe continuous growth models. 
This paper is structured as follows. 
In Sec.~\ref{sec_brief_review}, we present the one-parameter generalization of the logarithmic and exponential functions and a brief review of growth models.
In Sec.~\ref{sec:continuous_models}, we consider a simple type of Bernoulli's equation, which solution is the Zipf-Mandelbrot function. 
From the gradual generalization of the differential equations, their solutions retrieves the Richards' growth model~\cite{richards_1959}, Marusic and Bajzer model~\cite{marusic_1993} and finally Tsoularis and Wallace model~\cite{tsoularis_2002}. 
Since the Richards' model has an underlying microscopic model~\cite{idiart_2002}, one is able to interpret physically  the logaritmic function generalization parameter. 
Final remarks are presented in Sec.~\ref{sec:conclusion}. 

\section{Brief review}
\label{sec_brief_review}

To facilitate the connection between the one-parameter logarithmic and exponential generalizations and growth models, we briefly review these subjects in the following. 
We first present the simple geometrical arguments~\cite{arruda_2007} which we have used to obtain a one-parameter generalization of the logarithmic and exponential functions. 
This natural geometrical generalization coincides exactly to the one proposed by Tsallis~\cite{tsallis_qm} in the context of non-extensive thermodynamics~\cite{tsallis_1988}. 
Instead of calling the extra parameter as $q$, like in the precursor papers, we have used $\tilde{q} = 1 - q$, which is more convenient to stress the symmetry properties of these functions.
Next, we present some general continuous growth models and show how to retrieve the standard ones with the appropriate set of parameters.

\subsection{The $\tilde{q}$-logarithmic, and the $\tilde{q}$-exponential functions} 
\label{sec:geometric_generalization_log_exp_functions}

The $\tilde{q}$-logarithm function $\ln_{\tilde{q}}(x)$ is defined as the value of the area underneath the non-symmetrical hyperbole $f_{\tilde{q}}(t) = 1/t^{1-\tilde{q}}$ in the interval $t \in [1,x]$, with $x > 0$:
\begin{eqnarray}
\ln_{\tilde{q}}(x) & = & \int_1^x \frac{\mbox{d}t}{t^{1-\tilde{q}}} = \lim_{\tilde{q}' \rightarrow \tilde{q}}\frac{x^{\tilde{q}'} - 1}{\tilde{q}'} 
\; .
\label{eq:gen_log}
\end{eqnarray}
For any $\tilde{q}$, the area is negative for $0<x<1$, it  vanishes for $x=1$  [$\ln_{\tilde{q}}(1) = 0$], and it is positive for $x>1$. 
This function is \emph{not} the logarithmic function in the basis $\tilde{q}$ [namely $\log_{\tilde{q}}(x)$] but a one-parameter generalization of the natural logarithmic function definition. 

The $\tilde{q}$-exponential function $e_{\tilde{q}}(x)$ is defined as the $t$-value, in such a way that the area underneath $f_{\tilde{q}}(t) = 1/t^{1-\tilde{q}}$, in the interval $t \in [1,e_{\tilde{q}}(x)]$, is $x$.
This is the inverse of the $\tilde{q}$-logarithm function $e_{\tilde{q}}[\ln_{\tilde{q}}(x)] = x = \ln_{\tilde{q}}[e_{\tilde{q}}(x)]$ and it is given by: 
\begin{eqnarray}
e_{\tilde{q}}(x) & = & 
                               \lim_{\tilde{q}' \rightarrow \tilde{q}}[1 + \tilde{q}'x]_{+}^{1/\tilde{q}'} \; . 
\label{eq:q_exp}
\end{eqnarray}
Here the the operator $[a]_+ = \max(a,0)$ is necessary since $e_{\tilde{q}}(x)$ is not real if $\tilde{q} x < -1$. 
This is a non-negative function $e_{\tilde{q}}(x) \ge 0$ and $x = 0$ is a special point since $e_{\tilde{q}}(0) = 1$, independently of the $\tilde{q}$ value.

\subsection{Continuous growth models}

Population dynamics deals with the evolution of species interacting with themselves and the environment. 
The one-species (logistic-like) models are well described in terms of growth models. 
Essentially, growth models connect the \emph{relative growth rate} $d n(t)/[n(t)dt] = d \ln n(t)/dt$ of a population, with $n(t)$ individuals of a given species at time $t$, to an \emph{induced saturation function} $G_{r,n_{\infty}}(n)$:
$\mbox{d} \ln n / \mbox{d}t = G_{r,n_{\infty}}(n)$, where $n_{\infty} = n(\infty)$ is the \emph{carrying capacity} (environment saturation level),  which corresponds to the asymptotic value of the stationary solution [$\mbox{d} \ln n(t)/\mbox{d}t = G_{r,n_{\infty}}(n) = 0$] and $r$ is the \emph{intrinsic growth rate} (individual contribution). 
To compare populations in different environments, a better variable to work with is the ratio between the population size and the carrying capacity (final population size): $p(t) = n(t)/n_{\infty}$.
In this new variable, one has: 
\begin{equation}
\frac{\mbox{d} \ln p }{ \mbox{d}t} = G_{r}(p) \; ,
\label{eq:dp}
\end{equation}
where we notice that now, the induced saturation function depends only on $r$, which we write $G_r \propto r$.
Thus, the environmental contribution is taken into account into the variable itself.

In the following we consider one-species (logistic-like) models. 
The Malthus' (exponential) model is the simplest one [$G_{r}(p) = r$], which is extended basically to two other models:
\begin{itemize} 
\item the Verhulst (logistic) model [$G_{r}(p) = r (1 - p)$], whose discretization leads to the logistic map (prototype example in dynamical system) and 
\item the Gompertz's model [$G_{\kappa}(p) = - \kappa \ln p$], which is also used to calculate life insurances taxes~\cite{murray,boyce_diprima}. 
\end{itemize}

It is interesting to observe that in Ref.~\cite[p.~83]{montroll_west}, it is shown that the Verhuslt and Gompertz models can be  generalized by a simple function proposed by Montroll and Badger~\cite{badger_1974}, which in fact is exactly the Richards' model~\cite{richards_1959} 
\begin{equation}
G_{\kappa}(p) = \kappa \; \frac{1-p^{\tilde{q}}(t)}{\tilde{q}} \;, 
\end{equation}
where $\kappa = r \tilde{q}$.
When $\tilde{q} = 1$, one obtains the Verhust model and when $\tilde{q} \rightarrow 0$, one retrieves the Gompertz's model. 
It is interesting to point out that $\kappa$ and $r$ are not exactly the same intrinsic growth parameter, although they are linearly related. 

A more general description of growth models has been given recently by the Tsoularis and Wallace model~\cite[p.~22]{tsoularis_2002}. 
They considered the following induced saturation function:
\begin{equation}
G_{\kappa}(p) = \kappa \; p^{\alpha-1}(t) \; \left[ \frac{1-p^{\tilde{q}}(t)}{\tilde{q}} \right]^{\gamma} \; , 
\label{eq:tsoularis}
\end{equation} 
where $\kappa = r \tilde{q}^{\gamma} n^{\alpha-1}_{\infty}$. 
The formal solution of Eq.~(\ref{eq:dp}) with the above induced saturation function is given by:
\begin{equation}
B_{p^{\tilde{q}}(t)} \left(\frac{1-\alpha}{\tilde{q}},1-\gamma \right) - B_{p_0^{\tilde{q}}} \left(\frac{1-\alpha}{\tilde{q}},1-\gamma \right) = \kappa t \; ,
\label{eq:tsoularis_sol}
\end{equation}
where $B_x(a,b) = \int_0^{x} dt \; t^{a-1}(1 - t)^{b-1}$ is the \emph{incomplete beta function} and two sets  of values are acceptable: 
\begin{enumerate}
\item $\alpha < 1$, $\tilde{q} > 0$ and $\gamma < 1$, 
\item $\alpha > 1$, $\tilde{q} < 0$ and $\gamma < 1$. 
\end{enumerate}

Consider the special cases of the above model.
\begin{itemize}
\item For $\gamma = 0$, one has Zipf-Mandelbrot-like growth models, which has two special cases:
      \begin{itemize} 
      \item exponential, for $\alpha = 1$ and 
      \item power-law, for $\alpha > 1$.
      \end{itemize} 
\item For $\gamma = 1$, one retrieves the Marusic and Bajzer model~\cite{marusic_1993}, which generalizes other growth models:
\begin{itemize}
\item if $\tilde{q} = 1 - \alpha$, the generalized von Bertanffy's model or 
\item if $\alpha = 1$ and $\tilde{q} \ge -1$, the Richards' model.
\end{itemize}
\item For other $\gamma$ values, one obtains:
      \begin{itemize}
      \item For $\alpha = 1$ and $\tilde{q} \rightarrow 0$, the hyper-Gompertz's model $G_{\kappa}(p) = -\kappa \ln^{\gamma} p(t)$. 
      \item For $\tilde{q} = 1$, one obtains the Blumberg's growth model~\cite{blumberg_1968} used to model organ size evolution. 
The Blumberg's model reduces to the hyperbolic model for regenerative growth, if $\gamma = 1 + 1/N$ and $\alpha = 1 - 1/N$ (see Ref.~\cite{spencer_1966}).
            
      \item If $\alpha = 1+\tilde{q}(\gamma -1)$, Tsoularis and Wallace model reduces to the Turner \emph{et al.} model, which surprisingly admits analytical solution~\cite{turner_1976}.
      \end{itemize}
\end{itemize}
      
\section{$\tilde{q}$-generalized functions and growth models}
\label{sec:continuous_models}

In what follows we will first deal with a particular type of Bernoulli's equation 
$[\mbox{d}/\mbox{d}x  + p(x) ] y(x) = q(x) y^{1-\tilde{q}}(x)$, with $p(x) = 0$ and $q(x) = 1$. 
We show that this equation (kinetic equation of arbitrary order) is related to the hyper-Gompertz model. 
When a new linear term is added to this equation, it represents the Richard's growth models and in this case we have been able to give a physical interpretation to $\tilde{q}$.
If this new added term is non-linear, we obtain an equation, which represents the Marusic-Bajzer models. 
The effort term can be included considering the $\tilde{q}$-logarithm function raised to a power, which gives rises to the Tsoularis and Wallace model.  

\subsection{Zipf-Mandelbrot model}

The derivative of the $\tilde{q}$-exponential function with respect to $x$ is:
$\mbox{d} e_{\tilde{q}}(k x)/\mbox{d} x = k \; [e_{\tilde{q}}(k x)]^{1-\tilde{q}}$, so that the $\tilde{q}$-exponential function is the solution of the following non-linear first order differential equation~\cite{tsallis_bemski,montemurro_in_book,Niven_2006}:
\begin{equation}
\frac{\mbox{d} y(x)}{\mbox{d}x} = k \; y^{1-\tilde{q}}(x) \; , 
\label{eq:dif_q_eq}
\end{equation}
where $k$ is a constant. 
In Chemistry it is known as \emph{kinetic equation of order $1-\tilde{q}$}. 
This equation is not directly related to any growth model, but if one thinks that $y = \ln p$, then one has the hyper-Gompertz model. 

The solution of Eq.~(\ref{eq:dif_q_eq}) can be found rewriting it as~\cite{boyce_diprima}: $\tilde{q} y^{\tilde{q}-1} \, \mbox{d}y/\mbox{d}x = \tilde{q} k \Rightarrow \mbox{d}y^{\tilde{q}}(x)/\mbox{d}x = \tilde{q}k$, which is linear in the variable $v(x) = y^{\tilde{q}}(x)$ and can be solved: $v(x) = \tilde{q}kx+v_0$, where the ``initial'' condition is $v_0 = v(0) = y_0^{\tilde{q}}$ with $y_0 = y(0)$.
Thus, one writes:
\begin{equation}
\frac{y(x)}{y_0} = \left( 1 + \tilde{q} \; \frac{ x}{x_{\tilde{q}}} \right)^{1/\tilde{q}} = e_{\tilde{q}} \left( \frac{ x}{x_{\tilde{q}}} \right) \;,
\label{eq:sol_dif_q_eq}
\end{equation}
with $x_{\tilde{q}} = y_0^{\tilde{q}}/k$.

\subsection{Richards-like growth models}

Eq.~(\ref{eq:dif_q_eq}) can be generalized to take care of other interesting situations.
The following differential equation:
\begin{equation}
\frac{\mbox{d} y(x)}{\mbox{d}x} = m y(x) + (k - m) y^{1-\tilde{q}}(x) \; , 
\label{eq_bemski}
\end{equation}
has been used in growth models~\cite{marusic_1993} and other contexts~\cite{tsallis_bemski,montemurro_in_book}. 
Factoring $m y(x)$ and calling $\tilde{y} = [m/(m-k)]^{1/\tilde{q}}y(x)$, Eq.~(\ref{eq_bemski})  is written as:
\begin{equation}
\frac{\mbox{d} \ln \tilde{y}^{-1}(x)}{\mbox{d}x} = m \tilde{q} \, \ln_{\tilde{q}} \tilde{y}^{-1}(x) \; ,
\label{eq_bemski2}
\end{equation}
which leads to the Richard's model as we will show.

\subsubsection{Richards' growth model}

In Eq.~(\ref{eq_bemski2}), calling $t$ the independent variable $x$, $p = 1/\tilde{y}$ and $m \tilde{q} = - \kappa$, one rewrites it as:
\begin{equation}
\frac{\mbox{d} \ln p(t)}{\mbox{d}t} =  - \kappa \, \ln_{\tilde{q}} p(t) \; , 
\label{eq_bemski3}
\end{equation}
which solution is 
\begin{eqnarray}
p(t) & = & \frac{1}{e_{\tilde{q}}[\ln_{\tilde{q}}(p_0^{-1}) e^{- \kappa t}]} = e_{-\tilde{q}}[-\ln_{\tilde{q}}(p_0^{-1}) e^{- \kappa t}] 
\nonumber \\ 
& = & e_{-\tilde{q}}[\ln_{-\tilde{q}}(p_0) e^{- \kappa t}] \; .
\label{eq:solucao_modelo_generalizado}
\end{eqnarray}
This is the Richards' growth model~\cite{richards_1959}. 

The discretization of Eq.~\ref{eq_bemski3} leads to a generalization of the logistic map different from the previously considered one (Feigenbaum's map). 
An initial study of this recursive equation has been presented in Ref.~\cite{martinez:2008a}. 

\paragraph{Particular Cases.}

Some particular cases can of Eqs.~(\ref{eq_bemski3}) and~(\ref{eq:solucao_modelo_generalizado}) be retrieved:
\begin{itemize}
\item If $\tilde{q} = 1$, one recovers the Verhuslt's model (logistic equation):
\begin{equation}
\frac{\mbox{d} \ln p} {\mbox{d} t} = -\kappa \; \ln_1(p) = \kappa \; (1 - p) \; ,
\label{eq:verhuslt}
\end{equation}
which has the following solution:
\begin{eqnarray}
p(t)  & = & e_{-1}[(1 - p_0^{-1}) e^{- \kappa t}] = \frac{1}{1 + (p_0^{-1} - 1) e^{- \kappa t}} \; , 
\label{eq:logistic_equation}
\end{eqnarray}
where $p_0 = n_0/n_{\infty}$ with $n_0 = n(0)$ being the initial population size and $n(\infty) = n_{\infty}$ is the carrying capacity. 
\begin{itemize}
\item If $\kappa > 0$, then for $t \gg 1/\kappa$, $p(\infty) = 1$, i.e. the carrying capacity is an attractor. 
\item If $\kappa < 0$, the carrying capacity $n_{\infty}$ becomes a repulsor and the solutions  
$p(t)  =  1/[1 + (p_0^{-1} - 1) e^{|\kappa| t}] = e_{-1}[(1 - p_0^{-1}) e^{|\kappa| t}]$, either converge, for $n_0 < n_{\infty}$ (i. e. $p_0 < 1$) or diverge, otherwise. 
This divergence means that the population becomes non-limited for times greater than $t^{*} =  \ln(1 - p_0^{-1})/\kappa$, which is solution of $1 + (p_0^{-1} - 1) e^{\kappa t^{*}} = 0$ and depends on the initial value  $p_0$.
\item As $n_{\infty} \rightarrow \infty$, then $p \rightarrow 0$, and one has:
 \begin{equation}
 \frac{\mbox{d} \ln p}{\mbox{d}t} = \kappa \; ,
 \label{eq:malthus}
 \end{equation}
which is the Malthus' model.
The solution of Eq.~(\ref{eq:malthus}) is: 
\begin{equation}
p(t) = p_0 \; e^{\kappa t} \; .
\end{equation} 
\end{itemize}

\item If $\tilde{q} = 0$, one retrieves the Gompertz's model:
\begin{equation}
\frac{\mbox{d} \ln p}{\mbox{d} t} = - \kappa \; \ln p \; , 
\label{eq:gompertez_model}
\end{equation}
which solution is:
\begin{equation}
p(t) = p_0^{e^{-\kappa t}} \; .
\end{equation}
\item If $\tilde{q} = -1$, one has the monomolecular or Mitscherlich growth model~\cite[p.~33]{tsoularis_2002}:
\begin{equation}
\frac{d p(t)}{d t} = k [1 - p(t)] \; , 
\label{eq:Mitscherlich}
\end{equation}
so that
\begin{equation}
p(t) = 1 - (1-p_0) e^{-kt} \; .
\label{eq:mitscherlich}
\end{equation}
Observe that Mitscherlich model [Eq.~(\ref{eq:Mitscherlich})], which uses $dp/dt$ differs fundamentally from Verhulst model [Eq.~(\ref{eq:verhuslt})], which uses $d \ln p / dt$.

\end{itemize}

\paragraph{Microscopic Model.}

The competition between cell drive to replicate and inhibitory interactions, that are modeled by long range interaction among the cell, furnishes an interisting microscopic mechanism to obtain Richards' model~\cite{idiart_2002}. 
The long range interaction is dependend on the distance $r$ between two cells as a power law $r^{\gamma}$ and the cells have a fractal structure characterized by a fractal dimension $D_f$. 
Using Eq.~\ref{eq:gen_log}, one can write their Eq.~(7) of Ref.~\cite{idiart_2002} as: $\mbox{d}\ln n(t)/\mbox{d}t = \langle G \rangle - J \omega \ln_{\tilde{q}}[D_f n(t)/\omega]$, where $\omega$ is a constant related to geometry of the problem and $\tilde{q} = 1 - \gamma/D_f$. 
Here the parameter $\tilde{q}$ acquires a physical meaning related to the interaction range $\gamma$ and fractal dimension of the celular structure $D_f$. 
This physical interpretation of $\tilde{q}$ has only been possible due to Richards' model underlying microscopic description. 

Another point of view is to obtain particular cases from the microscopic model of Ref.~\cite{idiart_2002}.  
If all the cells interact with the same intensity  for a fixed $D_f \ne 0$, then $\gamma = 0$ and $\tilde{q} = 1$, so that Eqs.~(\ref{eq_bemski3}) and~(\ref{eq:solucao_modelo_generalizado}) lead to the Verhulst model. 
Another way to obtain the Verhulst model is to fix the cell interaction range $\gamma \ne 0$ and consider large fractal dimensionality $D_f \to \infty$. 
Both cases illustrates that the Verhulst model is a mean field model. 
If cell interaction range matches with the cell fractal structure then $\gamma = D_f$ and $\tilde{q} = 0$, so that Eqs.~(\ref{eq_bemski3}) and~(\ref{eq:solucao_modelo_generalizado}) lead to the Gompertz's model.
If $\gamma > D_f$, then $\tilde{q} < 0$ and Eq.~(\ref{eq:solucao_modelo_generalizado}) reads: $p(t) = 1 + \ln_{|\tilde{q}|}(p_0) e^{- \kappa t}$, for $\kappa t \gg 1$, which leads to exponential behavior. 
In particular if $\gamma = 2 D_f$, one retrieves the Mitscherlich model [Eq.~(\ref{eq:mitscherlich})].

\subsubsection{Richard-Schaefer's model}
\label{sec:schaefer}

If individuals can be inserted to or removed from a given population at fixed time intervals at a rate $\epsilon$, which is called the \emph{effort rate}, then one has the \emph{Schaefer's model}~\cite{murray,boyce_diprima}: 
\begin{equation}
\frac{\mbox{d} \ln p}{\mbox{d}t} = - \kappa \; \ln_{\tilde{q}}(p) - \epsilon \; .
\label{eq:schaerfer_gen}
\end{equation}
The solution of Eq.~(\ref{eq:schaerfer_gen}) is:
\begin{eqnarray}
\nonumber
p_{\epsilon,\kappa}(t) & = & \frac{e_{\tilde{q}}(\epsilon)}{e_{\tilde{q}}\{\ln_{\tilde{q}}[e_{\tilde{q}}(\epsilon) p_0^{-1}] e^{- (\kappa - \tilde{q}\epsilon) t}\}} \nonumber \\ 
                & = & e_{\tilde{q}}(\epsilon) \; e_{-\tilde{q}}\{-\ln_{\tilde{q}}[e_{\tilde{q}}(\epsilon) p_0^{-1}] e^{- (\kappa - \tilde{q}\epsilon) t}\} \; ,
\end{eqnarray}
so that when $\epsilon = 0$, one reobtains  Eq.~(\ref{eq:solucao_modelo_generalizado}).

\subsection{Marusic and Bajzer's growth model}

As proposed by M.~Marusic and Z.~Bajzer~\cite{marusic_1993} the equation
\begin{equation}
\frac{\mbox{d} y(x)}{\mbox{d} x} = a y^{\alpha}(x) + b y^{\beta}(x) \; ,
\label{eq:marusic}
\end{equation}
with $a,b, \alpha, \beta \in \Re$, furnishes the basis of the models of one-species dynamical population, which generalizes traditional growth models.

We remark that Eq.~(\ref{eq:marusic}) with $a>0$ and $b<0$ was first studied in Ref.~\cite{bertalanffy_1952}, which deals with quatitative laws in metabolism and growth.
A more fundamental derivation of Eq.~(\ref{eq:marusic}) has been furnished in Ref.~\cite{savageau_1979,savageau_1980} in the allometric morphogenesis (form or patern generation) context.
Marusic and Bajzer have considered arbitrary real parameters and solved in Eq.~(\ref{eq:marusic}). 

Further, we notice that one can rewrite Eq.~(\ref{eq:marusic}) as: $\mbox{d} y(x)/\mbox{d} x = a y^{\alpha}(x) + b y^{\alpha}(x)  (\beta - \alpha)[y^{\beta - \alpha}(x) -1 + 1]/(\beta - \alpha) = \tilde{a} y^{\alpha}(x) + \tilde{b} y^{\alpha}(x) \ln_{\beta - \alpha}(y)$, with $\tilde{a} = a + b$ and $\tilde{b} = b (\beta - \alpha)$. 
In this way Eq.~(2) in the Marusic and Bajzer paper~\cite{marusic_1993} is generalized and  retrived for the particular case $\beta \to \alpha$.

Calling $a = m$, $b = k - m$, $\alpha = 1 + \tilde{q}'$ and $\beta = 1 + \tilde{q}' - \tilde{q}$ in Eq.~(\ref{eq:marusic}),  one obtains:
\begin{equation}
\frac{\mbox{d} y(x)}{\mbox{d}x} = m y^{1 + \tilde{q}'}(x) +  (k - m) y^{1+ \tilde{q}'-\tilde{q}}(x) \; . 
\label{eq_bemski_g}
\end{equation}
Taking $\tilde{q}' = 0$, one recovers Eq.~(\ref{eq_bemski}).
Factoring $m y^{1 + \tilde{q}'}(x)$ and calling $\tilde{y} = [m/(m-k)]^{1/\tilde{q}}y(x)$, Eq.~(\ref{eq_bemski_g}) is written as:
\begin{equation}
\frac{\mbox{d} \ln_{\tilde{q}'} \tilde{y}^{-1}(x)}{\mbox{d}x} =  m \tilde{q} \left( \frac{k-m}{m} \right)^{\tilde{q}'/\tilde{q}} \, \ln_{\tilde{q}}[\tilde{y}^{-1}(x)] \; . 
\label{eq_bemski_g_2}
\end{equation}
Calling $t$ the independent variable $x$, $p = 1/\tilde{y}$ and $m \tilde{q} [(k-m)/m]^{\tilde{q}'/\tilde{q}} = \kappa$, one rewrites Eq.~(\ref{eq_bemski_g_2}) as:
\begin{equation}
\frac{\mbox{d} \ln_{\tilde{q}'} p(t)}{\mbox{d}t} =  - \kappa \, \ln_{\tilde{q}} p(t) \; . 
\label{eq_bemski_g_3}
\end{equation}

This equation has analytical solutions only for some particular values of $\tilde{q}'$.
For $\tilde{q}' = 0$, one recovers the Richards' model [Eq.~(\ref{eq_bemski3})].
For $\tilde{q} \ne 0$, other common growth models are obtained as limiting cases, as we see below.

\subsubsection{Generalized von Bertalanffy's model}

For $\tilde{q}' = \tilde{q}$, one has the generalized von Bertalanffy model~\cite[p.~47]{tsoularis_2002}, and Eq.~(\ref{eq_bemski_g_3}) has the following solution:
\begin{equation}
p(t) = e_{\tilde{q}}[\ln_{\tilde{q}}(p_0) \, e^{- \kappa t}] \; . 
\end{equation}
If $\tilde{q} = 1/3$, then $\tilde{q}' = \tilde{q} = 1/3$ and one recovers the specialized von Bertalanffy model~\cite{bertalanffy_1952}.

\subsubsection{Smith's model}

The Smith's model~\cite[p.~35]{tsoularis_2002} replaces the term $1-p$ in Verhulst model by $1 - (p + a \, dp/dt)$.
The resulting equation is
\begin{equation}
\frac{\mbox{d} \ln p(t)}{\mbox{d}t} = r \left[ \frac{1-p(t)}{1+ra \, p(t)} \right] \; , 
\end{equation}
which cannot be obtained immediately from Eq.~(\ref{eq_bemski_g_3}), but from an approximation of it by setting $\tilde{q} = 1$ and $\tilde{q}' = 1 - 0.473$.

\subsection{Tsoularis and Wallace's growth model}

The growth model proposed by A. Tsoularis and J. Wallace~\cite{tsoularis_2002} [given by Eq.~(\ref{eq:tsoularis})] can be obtained generalizing Eq.~(\ref{eq_bemski_g_3}) to: 
\begin{equation}
\frac{d \ln_{\tilde{q}'} p(t)}{dt} = \kappa \; \left[ - \; \ln_{\tilde{q}} p(t) \right]^{\gamma} \; . 
\label{eq_bemski_g_4}
\end{equation}
which solution is given by Eq.~(\ref{eq:tsoularis_sol}), with $\alpha = 1 - \tilde{q}'$.
For $\gamma = 1$, one retrieves the Marusic and Bajzer growth model [Eq.~(\ref{eq_bemski_g_3})]. 
For $\gamma \ne 1$, other common growth models are retrieved as limiting cases.

The parameter $\gamma$ has the effect of an effort rate.
This can be seen rewriting Eq.~(\ref{eq_bemski_g_4}) in terms of a supplementary $\tilde{q}$-logarithmic function:
\begin{equation}
\frac{d \ln_{\tilde{q}'} p(t)}{dt} = \tilde{\kappa} \; \ln_{\gamma} \left[ - \; \ln_{\tilde{q}} p(t) \right] - \tilde{\varepsilon} \; , 
\label{eq_bemski_g_5}
\end{equation}
where $\tilde{\kappa}= \kappa \gamma$ and $\tilde{\varepsilon} = - \kappa$.
For comparison, see Eq.~(\ref{eq:schaerfer_gen}).

\subsubsection{Hyper-Gompertz's model (or Exponential polynomial)}

For $\tilde{q}' = \tilde{q} = 0$ in Eq.~(\ref{eq_bemski_g_4}), one finds the hyper-Gompertz's model
\begin{equation}
\frac{d \ln p}{d t} = \kappa \; [- \ln p]^{\gamma} \; . 
\label{eq:gompertz_gen} 
\end{equation}
Calling $y = \ln p$, this equation becomes Eq.~(\ref{eq:dif_q_eq}), whose solution is given by Eq.~(\ref{eq:sol_dif_q_eq}).  
In the original variable, the solution of Eq.~(\ref{eq:gompertz_gen}) is:
\begin{eqnarray}
p(t) & = & p_0^{e_{1-\gamma}[-\kappa t/(\ln p_0)^{1-\gamma}]} 
       =  e^{\ln (p_0) e_{1-\gamma}[-\kappa t/(\ln p_0)^{1-\gamma}]}\; ,
\end{eqnarray}
where we point out the presence of a generalized exponential function in the argument of an ordinary one. 

\subsubsection{Blumberg's model}

For $\tilde{q} = 1$ in Eq.~(\ref{eq_bemski_g_4}), one retrieves the Blumberg model~\cite{blumberg_1968}, which has been proposed to model organ size evolution:
\begin{eqnarray}
\frac{d \ln_{\tilde{q}'} p}{dt} & = & \kappa \; [-\ln_{1}(p)]^{\gamma} \nonumber \\ 
\frac{d p}{dt}                         & = & \kappa  p^{1- \tilde{q}'} (1-p)^{\gamma} \; ,
\label{eq:blumberg_model}
\end{eqnarray}
which can be rewritten as: 
\begin{equation}
\int_{p_0}^{p(t)} dp \; p^{\tilde{q}'-1}(1 - p)^{-\gamma} = \kappa t \; .
\end{equation}
If $\tilde{q}' > 0$ and $\gamma < 1$, one has a closed analytical solution [as in Eq.~(\ref{eq:tsoularis_sol}), $\alpha = 1 - \tilde{q}'$]: 
\begin{equation}
B_{p(t)}(\tilde{q}', 1 - \gamma) - B_{p_0}(\tilde{q}', 1 - \gamma) = \kappa t \;.
\end{equation}

Notice that in Eq.~(\ref{eq:blumberg_model}) for: 
\begin{itemize}
\item $\gamma = 0$, one has the power law induced saturation function $G_{\kappa}(p) = \kappa p^{1-\tilde{q}'}$ and 
\item $\gamma = 1 + 1/N$ and $\tilde{q} = 1 - 1/N$, one has the \emph{hyperbolic model} for regenerative growth (see Ref.~\cite{spencer_1966}). 
\end{itemize}

\subsubsection{Turner \emph{et al.} model (Generic growth)}

For $\tilde{q}' = \tilde{q}(\gamma-1)$ in Eq.~(\ref{eq_bemski_g_4}), one retrieves the Turner \emph{et al.} model~\cite{turner_1976}:  
\begin{eqnarray}
\frac{d \ln p(t)} {d t} & = & \kappa \; p^{\tilde{q} (1-\gamma)}(t) \; \left[ - \ln_{\tilde{q}}p(t) \right]^{\gamma}  \; ,
\end{eqnarray}
which has the following analytical solution:
\begin{eqnarray}
p(t) 
     & = & e_{-\tilde{q}} \left\{ \ln_{-\tilde{q}}(p_0) \; e_{1-\gamma} 
                          \left[ -\kappa t (\ln_q p_0^{-1})^{\gamma-1} 
                         \right] 
                         \right\} \; .
\end{eqnarray}
 
The Turner \emph{et al.} model can be solved analytically only because of the judicious choice of $\tilde{q}'$. 

\subsubsection{Retrieving growth models}

\begin{figure}[ht]
\begin{center}
\includegraphics[angle = -90, width = \columnwidth]{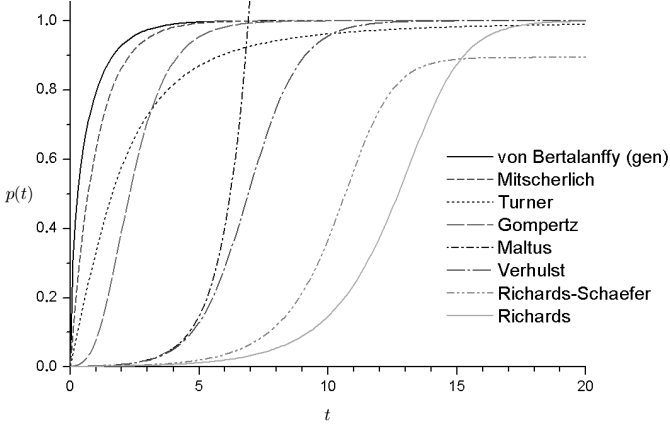}
\caption{Models which present analytical solution from Eq.~(\ref{eq_bemski_g_4}).
The curves have been obtained with the following values: $p_0 = 0.001$, $\alpha = 0.1$ and $\epsilon = -0.1$.
For parameters of arbitrary value in Tab.~\ref{Tab:Resumo}, we have taken: $\kappa = 1$, $q = 2$, and $\gamma = 1.5$.}
\label{fig:resumo}
\end{center}
\end{figure}

As one can see, all continuous growth models considered in this paper are generalized by Eq.~(\ref{eq_bemski_g_4}).
To recover those models, take appropriate values of $\tilde{q}'$, $\tilde{q}$, $\gamma$ and $\kappa$ shown in Tab.~\ref{Tab:Resumo}.
Some particular cases have analytical solution and they are shown in Fig.~\ref{fig:resumo}.
It can be observed that for the Richard-Schaefer model, the asymptotic value is $p_{\infty} = e_{\tilde{q}}(\epsilon) = \sqrt{0.8} = 0.89 \ldots $, due to the effort $\epsilon = -0.1$ rate, instead of $p_{\infty} = 1$.

\begin{table*}[htb]
\begin{center}
\begin{tabular}{l|c|c|c|c|l}
\hline
Model                        &            $\tilde{q}'$ &  $\tilde{q}$ & $\gamma$ & $\kappa$ & Equation \\
\hline
Malthus (exponential)        &                       0 &       *      &        0 &           $r$ & $d \ln p/dt = r$ \\
Verhulst (logistic)          &                       0 &            1 &        1 &           $r$ & $d \ln p/dt = r (1-p)$ \\
Gompertz                     &                       0 &            0 &        1 &             * & $d \ln p/dt = -\kappa \ln p$ \\
hyper-Gompertz         &                       0 &            0 &        * &             * & $d \ln p/dt = \kappa (-\ln p)^\gamma$ \\
Richards                     &                       0 &            * &        1 & $r \tilde{q}$ & $d \ln p/dt = r (1-p^{\tilde{q}})$ \\
Tsoularis and Wallace        &                       * &            * &        * &             * & $d \ln_{\tilde{q}'} p/dt = \kappa (-\ln_{\tilde{q}} p)^\gamma$ \\
Marusic and Bajzer           &                       * &            * &        1 &             * & $d \ln_{\tilde{q}'} p/dt = -\kappa \ln_{\tilde{q}} p$ \\
Mitscherlich (monomolecular) &                       1 &            1 &        1 &             * & $\ln_1 dp/dt = dp/dt = \kappa (1-p)$ \\
Blumberg                     &                       * &            1 &        * &             * & $d \ln_{\tilde{q}'} p/dt = \kappa (1-p)^\gamma$ \\
Turner \emph{et al.}                      & $1+\tilde{q}(1-\gamma)$ &            * &        * &             * & $d \ln p/dt = \kappa \, p^{\tilde{q} (1-\gamma)} (-\ln_{\tilde{q}}p)^{\gamma}$ \\
Specialized von Bertalanffy  &                     1/3 &          1/3 &        1 &             * & $d \ln_{1/3} p/dt = - \kappa \ln_{1/3} p$ \\
Generalized von Bertalanffy  &             $\tilde{q}$ &            * &        1 &             * & $d \ln_{\tilde{q}} p/dt = - \kappa \ln_{\tilde{q}} p$ \\
Smith                        &               $1-0.473$ &            1 &        1 &             * & approximation \\
\hline
\end{tabular}
\end{center}
\caption{Growth models which can be obtained from Eq.~(\ref{eq_bemski_g_4}). 
The asterisk means that the considered parameter can take an arbitrary value within its domain.}
\label{Tab:Resumo}
\end{table*}

\section{Conclusion}
\label{sec:conclusion}

The main objective of this paper is to show that the one-parameter generalization of the logarithmic and exponential function obtained from non-extensive statistical physics is suitable to describe a large class of continuous population dynamics (growth) model. 
On one hand, the $\tilde{q}$-logarithmic function appears in the differential equations that define the models.
On the other hand, the $\tilde{q}$-exponential function appears in the solutions of those differential equations, when it is analytically available. 
In the presentation we have gradually increased the number of parameters of the differential equation to establish connection with growth models. 
Eq.~(\ref{eq_bemski_g_4}) represents the most general model, the one of Tsoularis and Wallace. 
It has a simple form when written in terms of $\tilde{q}$-logarithmic function. 
A physical interpretation has been given to $\tilde{q}$ in the Richards' model. 
This has been possible due to the microscopic model presented in Ref.~\cite{idiart_2002}.

\section*{Acknowledgments}

The authors thank N. A. Alves, A. L. Esp\'{\i}ndola and R. da Silva for fruitful discussions.
We also thank J. Arenzon for calling our attention to Ref.~\cite{idiart_2002}. 
ASM acknowledges the Brazilian agencies CNPq (305527/2004-5, 303990/2007-4 and 476862/2007-8) and FAPESP (2005/02408-0) for support. 
RSG also acknowledge CNPq (140420/2007-0) for support.


\end{document}